\documentclass[aps, preprint, superscriptaddress, amsmath,amssymb]{revtex4}

\usepackage{txfonts}
\usepackage{graphicx}
\usepackage{color}
\usepackage[sort&compress]{natbib}
\begin{document}


\title{Wide angle and polarization independent chiral metamaterial absorber}

\author{Bingnan Wang}
\affiliation{%
Ames Laboratory and Department of Physics and Astronomy, Iowa State University,
Ames, Iowa 50011, USA
}%

\author{Thomas Koschny}%
\affiliation{%
Ames Laboratory and Department of Physics and Astronomy, Iowa State University,
Ames, Iowa 50011, USA
}%

\affiliation{%
Institute of Electronic Structure and Laser, FORTH, and Department of Materials
Science and Technology, University of Crete, 71110 Heraklion, Crete, Greece
}%

\author{Costas M. Soukoulis}%
 \email{soukoulis@ameslab.gov}
\affiliation{%
Ames Laboratory and Department of Physics and Astronomy, Iowa State University,
Ames, Iowa 50011, USA
}%
\affiliation{%
Institute of Electronic Structure and Laser, FORTH, and Department of Materials
Science and Technology, University of Crete, 71110 Heraklion, Crete, Greece
}%

\begin{abstract}
We propose a resonant microwave absorber based on a chiral metamaterial. We show, with both numerical simulations and experimental measurements, that the absorber works well for a very wide range of incident angles for different polarizations. The proposed absorber has a compact size and the absorption is close to one for a wide range of incident angles and it is a good candidate for potential applications.
\end{abstract}


 \maketitle

The research in electromagnetic metamaterials has been growing rapidly in recent years. With metamaterials, peculiar behaviors such as negative refraction, which are not seen in natural materials, can be obtained. Potential applications, such as super-lensing with a flat slab, cloaking, miniature antennas, as well as many other devices have been proposed and studied, covering the electromagnetic spectrum from microwave to visible regime (for reviews of the metamaterial field, see~\cite{MMandNRIreview.Smith.Science.2004,NIM.Soukoulis.AdvMatt.2007, NRIoptical.Soukoulis.Science.2007}). Most of the metamaterials are made of periodically arranged metallic structures much smaller than the working wavelength in size. To characterize the electromagnetic responses the metamaterials are treated as homogeneous media such that macroscopic parameters, electrical permittivity $\epsilon$ and magnetic permeability $\mu$, can be assigned~\cite{SRR.Pendry.IEEE.1999}. The major advantage of metamaterials over natural materials is that the macroscopic parameters can be designed to have desired values. Negative refraction, which is the primary goal of metamaterial research, can be achieved when both $\epsilon$ and $\mu$ are negative~\cite{Veselago}. Structures with electric resonances and magnetic resonances are utilized to fulfill this purpose. Since most proposed metamaterials are metallic resonant structures and rely on strong resonances, losses are inevitable. The existence of losses deteriorates the performance of potential devices such as superlenses~\cite{Pendry.Superlens.PRL.2000}. Different ways to reduce the losses have been studied, including the use of low loss materials, the optimization of designs~\cite{Loss.Jiangfeng.OE.2008} and the use of gain materials to compensate losses~\cite{Gain.Wegener.Soukoulis.OE.2008}. 

Instead of trying to reduce the losses, very recently, ideas have been proposed to build resonant absorbers with metamaterials~\cite{Absorber.PRL.100.207402.2008, Absorber.OE.2008, Absorber.PRB.78.241103.2008, Absorber.PRB.79.045131.2009}, as well as other metallic nanostructures~\cite{Absorber.MarcusPhC.PRB.2009}.  The absorption is defined as $A(\omega)=1-R(\omega)-T(\omega)$, where $A(\omega)$, $R(\omega)$ and $T(\omega)$ are the absorption, the reflection, and the transmission as functions of frequency $\omega$, respectively. It is straightforward to get the two design principles to make the absorption as close to unity as possible: minimize $R(\omega)$ and minimize $T(\omega)$. To minimize the reflection, we can tune the parameters of the metamaterial to get the effective $\epsilon$ and $\mu$ matched so that the impedance of the metamaterial $z=\sqrt{\mu/\epsilon}$ is equal to 1 and matched to the free space. So the reflection can in principle be eliminated. To minimize the transmission, the metamaterial needs to be designed so that the imaginary parts of $\epsilon$ and $\mu$ are as large as possible since they correspond to the loss in the metamaterial. Although it is difficult to get the transmission eliminated by one single layer of the metamaterial, there are ways to achieve unit absorption. The first is to use multiple layers of such metamaterial films to eliminate the transmission~\cite{Absorber.PRL.100.207402.2008,  Absorber.PRB.79.045131.2009}; the second is to use a ground plane to reflect the transmitted wave back~\cite{Absorber.PRB.78.241103.2008, Absorber.MarcusPhC.PRB.2009}. These two approaches have their advantages and disadvantages. The first approach can obtain in principle perfect absorption at the resonance peak, with the sacrifice of increasing thickness. The second approach can give a very thin absorber, but the absorption may not be perfect.

Most of the proposed metamaterial absorbers are composed of conducting electric resonators on two sides of a dielectric substrate~\cite{Absorber.PRL.100.207402.2008, Absorber.OE.2008, Absorber.PRB.79.045131.2009}. The electric response can be obtained from the excitation of the electric resonators by the electric field, and the magnetic response is provided by the anti-parallel currents on the two sides of the substrate~\cite{Zhou.ShortWirePairs.PRB.2006, Dolling.Soukoulis.NIM780nm.OL.2007}. These absorbers depend strongly on the polarization of the incident waves~\cite{Absorber.PRL.100.207402.2008, Absorber.OE.2008} as well as the incident angle. They work only for one polarization at normal incidence and the absorption drops rapidly for off-normal incidence cases. The absorber in Ref.~\cite{Absorber.PRB.79.045131.2009}, which is composed of paired metallic rods symmetric on two sides of a substrate, covers a wider angle but only works for one polarization. The design of periodic metallic strips with a ground plate~\cite{Absorber.MarcusPhC.PRB.2009} can have an absorption of over $80\%$ at $70^{\circ}$, but is still limited to one particular polarization. In Ref.~\cite{Absorber.PRB.78.241103.2008}, the metamaterial absorber, composed of split-ring shaped electric resonators, operates over a wide range of incidence angles for both transverse electric (TE) and transverse magnetic (TM) waves. However, the electric resonators in the absorber is not symmetric and still have azimuthal dependence.

In this brief report, we propose a new type of resonant absorbers that are made of chiral metamaterials. The chiral metamaterial absorber is shown, by both numerical simulations and experimental measurements, to be angle and polarization independent. Near perfect absorption can be achieved at the resonance. Moreover, the proposed absorber has a thickness of merely $1/5$ of the working wavelength $\lambda$. Although the modeling and experiments are done in microwave frequencies, the absorber can be easily scaled to find applications at higher frequency regimes.

Chiral metamaterials have been proposed as an alternative route to get negative refraction~\cite{ChiralNihility.Tretyakov.JElecWave.17.695.2003, Pendry.Science.Chiral}. Composed of chiral resonators, cross-coupling between the magnetic and electric fields happens at the resonance.  The cross-coupling effect, characterized by the dimensionless chirality parameter $\kappa$, is generally very small ($|\kappa| \ll$ 1) in natural chiral molecules. In chiral metamaterials, the chiral resonators can be designed to have strong resonances and have a much larger $\kappa$. The refractive indices of right-handed circular polarization $n_+$ and left-handed circular polarization $n_-$, which are the two eigen-solutions of electromagnetic waves in chiral media, are differentiated due to $\kappa$. $n_{\pm} = n \pm \kappa$, where  $n=\sqrt{ \epsilon \mu}$~\cite{EMWaveBImedia.1994}. When $\kappa$ is strong enough, one index becomes negative. Very recently, negative refraction by chiral metamaterials is demonstrated by experiments with different designs~\cite{Chiral.Pulm.PRB.2008, Zhang.NegIndChiralMM.PRL.102.023901.2009, Crosswire.Zhou.PRB.2009, ChiralSRR.Wang.draft.2009}. Similar to conventional metamaterials, chiral metamaterials require strong resonances to get a large $\kappa$, thus losses are also associated with chiral metamaterials. In microwave experiments, the major loss is shown to be dielectric loss due to the lossy substrates used. The loss can be reduced significantly by using low-loss substrates. On the other hand, resonant absorbers can be made with chiral metamaterials.

\begin{figure}
\centering
\includegraphics[width=0.9\textwidth]{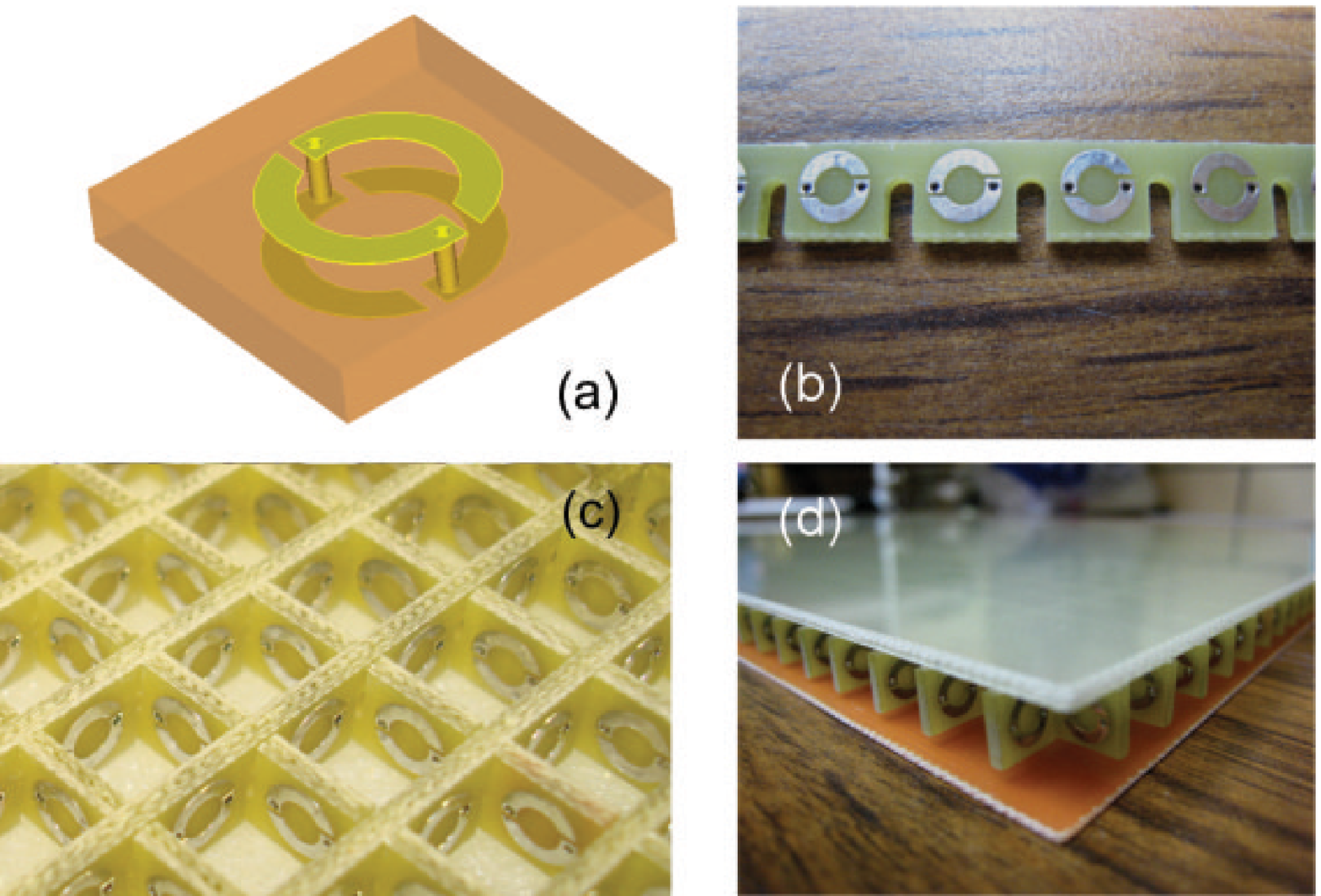}
    \caption{(color online) (a) The structure of the chiral SRR. (b) The chiral SRRs are fabricated on PCBs and cut into strips. The PCBs have a dielectric constant of $\epsilon_r = 3.76$ with loss tangent $0.0186$ and thickness of $1.6$ mm. The metal structures are built on copper with a thickness of $36 \, \mu$m. The SRRs are 2-gap split rings with an identical gap width of $0.3$ mm. The inner radius of the rings is $1.25$ mm and the outer radius is $2.25$ mm. The distance between adjacent rings is $8$ mm. (c) The strips are then interlocked to form the chiral metamaterial slab. (d) The metal ground plate and the cover plate are attached to the metamaterial slab to obtain a metamaterial absorber. The metal ground plate is also a PCB with copper on one side. The cover plate is a PCB with no copper cover. } \label{fig:structure}
\end{figure}

The chiral metamaterial in this study is based on Ref.~\cite{ChiralSRR.Wang.draft.2009}.  First proposed to develop three-dimensional chiral metamaterials~\cite{Marques.PRB.76.245115.2007, Marques.MicroOptTechLett.49.2606.2007, Marques.PRB.77.205110.2008}, the elemental structure is formed by two identical split-ring resonators (SRRs) separated by a dielectric substrate, and interconnected by vias, as shown in Fig.~\ref{fig:structure}(a). This is a chiral version of SRRs. To build a chiral metamaterial, the chiral SRRs are fabricated on printed circuit boards (PCBs) in arrays, and cut into long strips with grooves between neighboring cells, see Fig.~\ref{fig:structure}(b). The strips are then interlocked, with the guidance of the grooves, to form a metamaterial slab with square cells, with the chiral SRRs on the walls of each cell (Fig.~\ref{fig:structure}(c)). The chiral metamaterial has been shown to have negative refraction due to strong chirality at the resonance frequency~\cite{ChiralSRR.Wang.draft.2009}. Meanwhile, significant loss is associated with the resonance, which, shown by numerical simulations, is due mainly to dielectric loss in the lossy FR-4 board. The loss can thus be reduced by choosing substrates with a smaller dissipation factor. On the other hand, a metamaterial absorber can be made out of the current structure. 

A chiral metamaterial absorber can be build from the chiral metamaterial slab, backed with a ground copper plate, and covered with a dielectric plate, see Fig.~\ref{fig:structure}(d). For normal incidence, the incident wave is rotated by an angle $\theta$ when it reaches the ground plane; when the wave is reflected back to the chiral metamaterial, it is rotated by the same angle $\theta$, but in the opposite direction. Therefore, the polarization is preserved after the reflection, due to the reciprocity of the metamaterial. In the case of oblique incidence, part of the reflected wave is transformed. Simulations have shown that the polarization transformation is less than $3\%$ even the incident angle is $85^{\circ}$ off-normal.

\begin{figure}
\begin{minipage}[b]{1\linewidth}
\centering
\begin{tabular}{cc}
\includegraphics[width=0.45\textwidth]{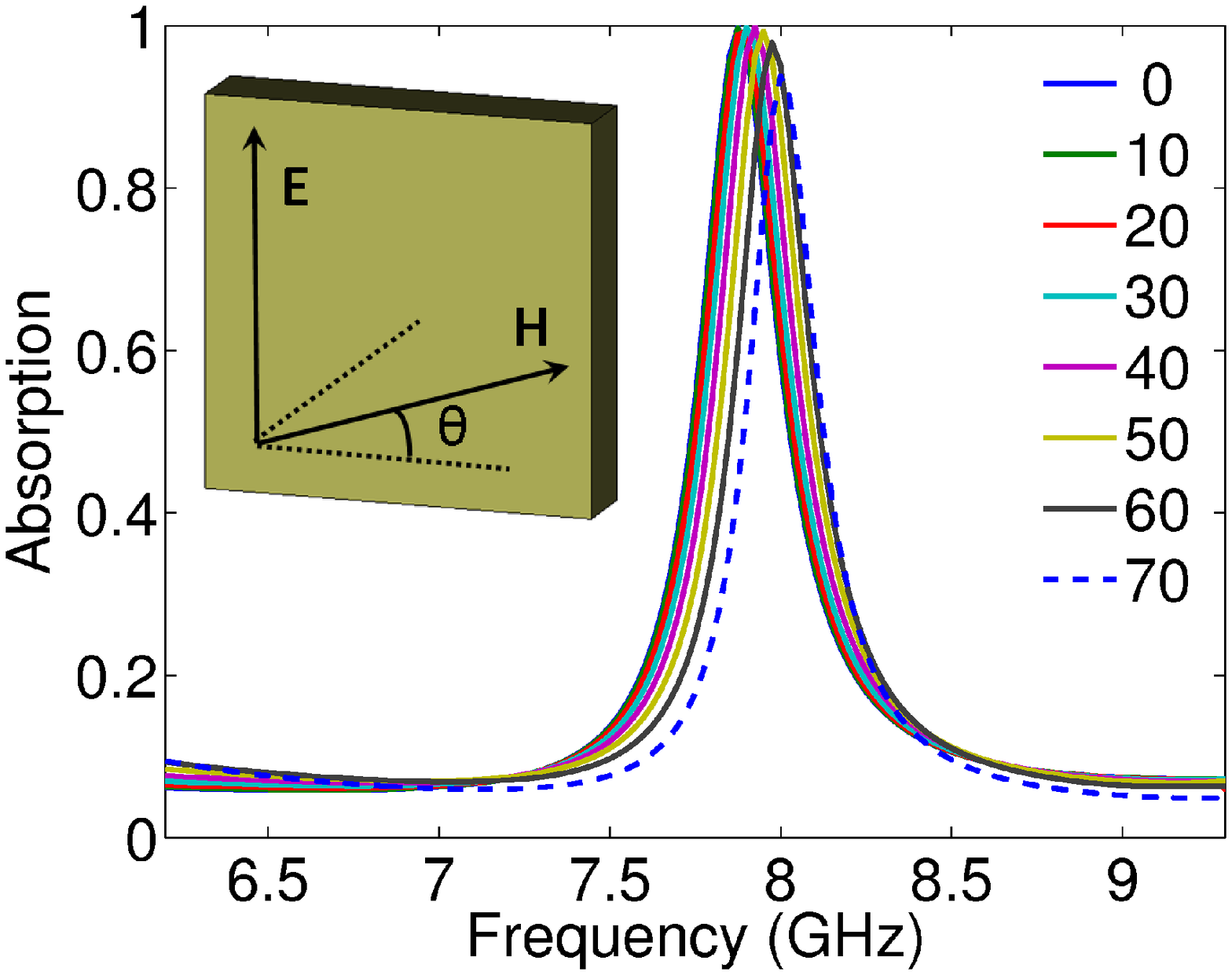}&
\includegraphics[width=0.45\textwidth]{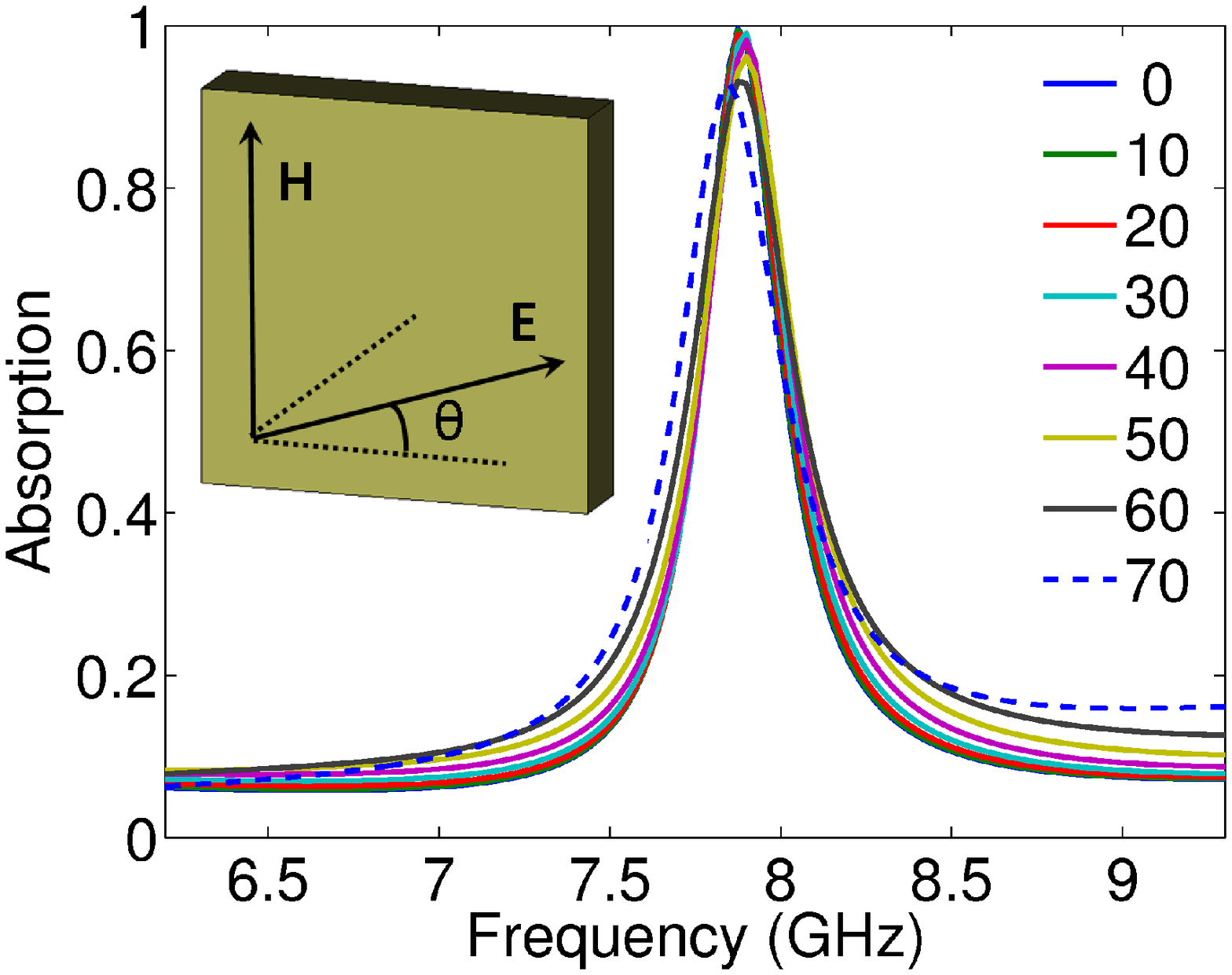}\\
(a)&(b)\\
\end{tabular}
\end{minipage}
\caption{(color online) The simulation results of absorption at different angles $\theta$ for (a) TE polarization and (b) TM polarization. Insets illustrate the two polarizations and the angle $\theta$.}\label{fig:simulation}
\end{figure}

The numerical simulations are done in CST Microwave studio. With the help of unit cell boundaries, an infinitely large slab of the absorber is simulated. With the ground plane in the back, the transmission is eliminated and the absorption is calculated by $A_{E}=1-R_{EE}-R_{HE}$ for TE polarization and $A_{H}=1-R_{HH}-R_{EH}$ for TM polarization, where the subscript $E$ indicates TE polarization and $H$ indicates TM polarization. The term $R_{EE}$ means the reflection coefficient of a TE wave from a incident TE wave and $R_{HE}$ means the reflection coefficient of a TM wave from a incident TE wave.  The absorption is calculated for different incident angles, as shown in Fig.~\ref{fig:simulation}(a) for TE polarization and Fig.~\ref{fig:simulation}(b) for TM polarization. For TE polarization, the absorption is almost unity at normal incidence and remains above $98\%$ until $60^{\circ}$. The peak absorption drops for large incident angles but is still above $90\%$ even when the incidence angle is $70^{\circ}$. When taking into account the small shift of the center frequency ($1.25\%$ from $0^{\circ}$ to $70^{\circ}$), the absorption at the peak frequency of normal incidence case is still more than $40\%$ ($60\%$) at $70^{\circ}$ ($60^{\circ}$). For TM polarization, the absorption is above $90\%$ for all incident angles and the center frequency shift is less than $1\%$. Moreover, since the metamaterial slab is uniaxial in plane, the absorber has no azimuthal dependence for either polarization. All these results show that the proposed design can be used as a perfect absorber.

\begin{figure}
\begin{minipage}[b]{1\linewidth}
\centering
\begin{tabular}{cc}
\includegraphics[width=0.45\textwidth]{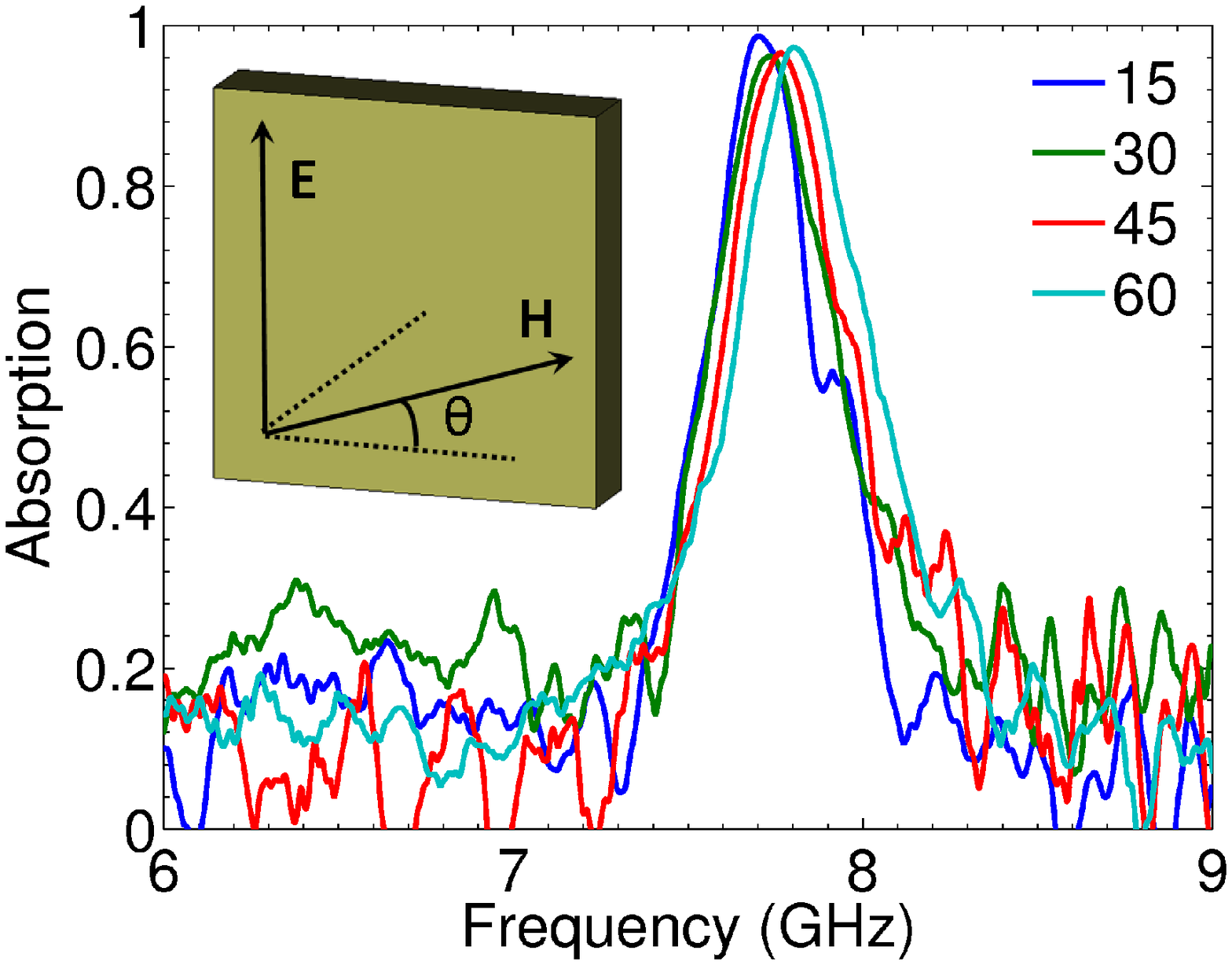}&
\includegraphics[width=0.45\textwidth]{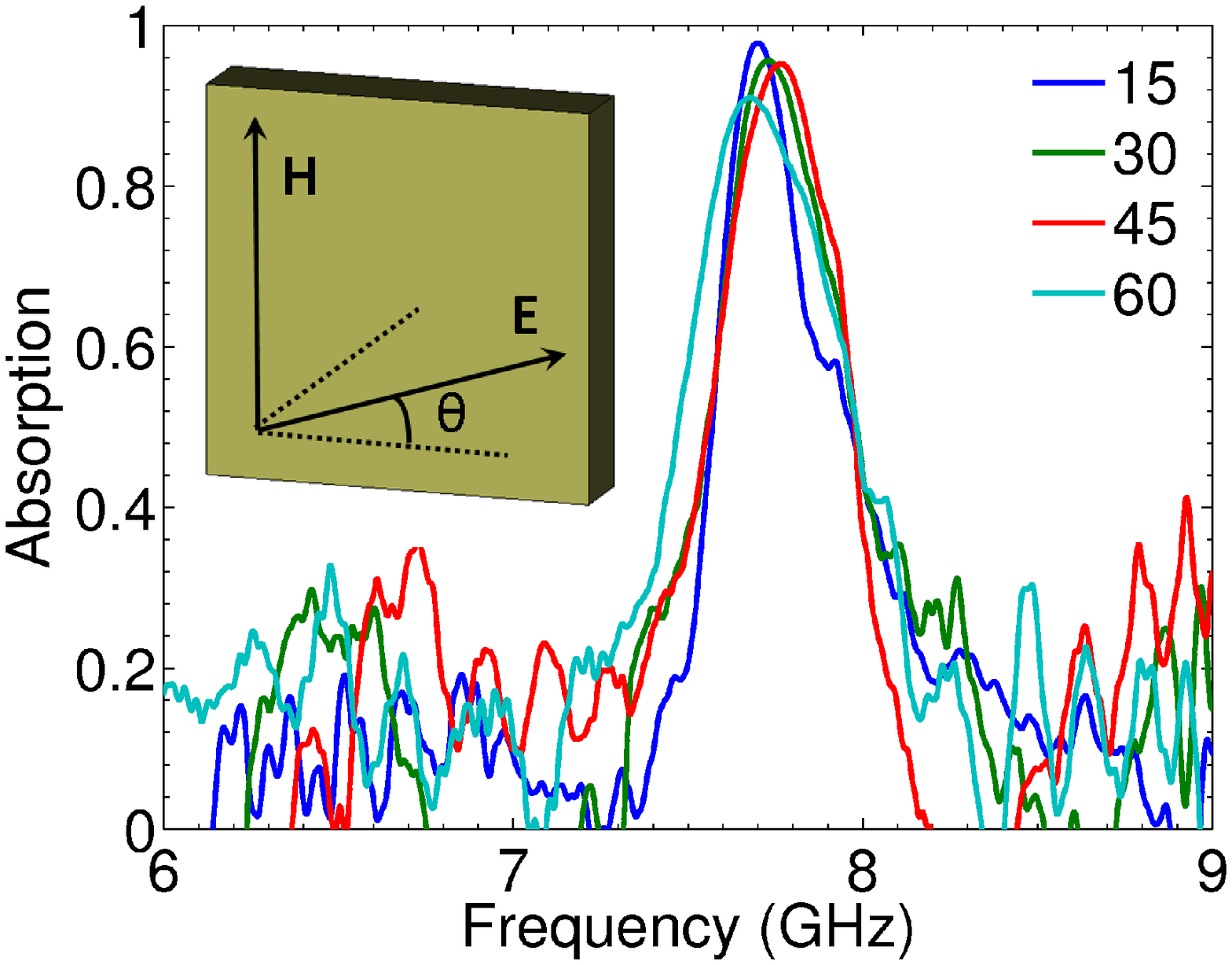}\\
(a)&(b)\\
\end{tabular}
\end{minipage}
\caption{(color online) The experiment results of absorption at different angles $\theta$ for (a) TE polarization and (b) TM polarization. Insets illustrate the two polarizations and the angle $\theta$.}\label{fig:experiment}
\end{figure}

Experiments have also been done to test the absorption behavior of the fabricated sample. A vector network analyzer (Agilent E8364B) and a pair of standard gain horn antennas are used to measure the reflection coefficient from the sample. The EM wave from the horn antenna is linearly polarized. By changing the orientation and the angle of the antennas, the reflection coefficient for both TE and TM polarizations at different incident angles can be measured. The reflection from a metal plate without the absorber is used for normalization. The absorption is then calculated and shown in Fig.~\ref{fig:experiment}(a) and Fig.~\ref{fig:experiment}(b). The absorption peaks at resonance are smaller than the simulation results. This is partly due to the fact that the actual dissipation factor of the PC boards may be different from the one used in simulations. The fabrication imperfection contributes to the  discrepancies as well. The measured absorption off resonance is generally higher than the simulation results and has many small oscillations. This is caused by the scattering from imperfections in the fabricated structure and the scattering between the horn antennas due to their relatively large aperture.  In spite of these discrepancies, the experimental measurements agree with the simulation results that high absorption can be obtained at the resonance for a wide range of incident angles for different polarizations. With these nice absorption properties, the proposed metamaterial absorber is still relatively thin, with a thickness of $1/5$ of the working wavelength.

The nature of the resonance and the absorption can be better understood by the surface current density distribution~\cite{ChiralSRR.Wang.draft.2009}. When the structure is excited by external field, both magnetic dipole and electric dipole exist and they are both parallel to the axis of the SRRs. While the induced current around in the SRRs gives the magnetic dipole, electric charges accumulate on the two SRRs with opposite signs, which introduce a strong electric field between the top and bottom SRRs and give the electric dipole in the same direction as the magnetic dipole. An electric dipole can excite a magnetic dipole in a similar way. Due to this cross-coupling effect and the nonplanar arrangement of the resonators, the resonance can be excited by different polarizations at a wide range of incident angles.

The idea to fabricate microwave absorbers with chiral media can be dated back to 1980's~\cite{Absorber.Chiral.1987, Absorber.Chiral.Engheta.1989}. The interest is due to the extra macroscopic parameter $\kappa$ other than $\epsilon$ and $\mu$. The authors~\cite{Absorber.Chiral.1987, Absorber.Chiral.Engheta.1989} believed that $\kappa$ can give extra flexibility in the design of phase-matched microwave absorbers and thus provide improvement to the performance. A debate started soon after these publications, concerning whether $\kappa$ can really affect the impedance and enhance the absorption~\cite{Absorber.Chiral.Comment.1989, Absorber.Chiral.Argue.1992}. Meanwhile, there is not much experimental support for the superiority of chiral absorbers. A discussion at the conference Bianisotropics'97 concluded that chirality dos not lead to superior absorption~\cite{Absorber.Chiral.Report.1997}. Later experiment results with chiral and achiral resonant structures show that, resonance structures, no matter chiral or not, in a lossy host can significantly enhance the absorption at the resonance frequency and chirality does not play a role in the enhancement of absorption~\cite{Absorber.ChiralResonance.2001}. The superior absorption properties of our design are due to the very strong resonance that can be excited by both electric and magnetic fields, and the compact size of the resonators.

In summary, the absorption properties of a resonant absorber made from  chiral metamaterials are studied. The absorber is shown, by both numerical simulations and experimental measurements, to be working  for a very wide angle. The absorber is also independent of polarizations and azimuthal angles. Due to the compact size of the resonators, the proposed absorber has a very small thickness (about $1/5$ of the center wavelength). All those features make the proposed design a perfect absorber. By scaling the parameters, the design in microwave can be applied to higher frequency regimes.

Work at Ames Laboratory was supported by the Department of Energy (Basic Energy
Sciences) under contract No. DE-AC02-07CH11358. This work was partially
supported by AFOSR under MURI grant (FA 9550-06-1-0337).

\clearpage

\end{document}